# ENHANCED OSL EMISSION FROM α- $Al_2O_3$ PRODUCED IN THE PRESENCE OF HALLOYSITE NANOCRYSTALS

C. B. B. M. FERREIRA and E. J. GUIDELLI

Departamento de Física - FFCLRP- Universidade de São Paulo, Ribeirão Preto, Brasil, SP 14040-901

Corresponding author: Tel: +55 (16) 3315-0080

E-mail address: guidelli@usp.br (E. J. Guidelli)

## Abstract

This paper developed a seed-mediated synthesis of α- $Al_2O_3$ by the combustion method using halloysite nanocrystals as seeds and analyzed the dosimetric characteristics of these samples, including radioluminescence (RL), thermoluminescence (TL), and optically stimulated luminescence (OSL). Both SEM images and XRD pointed to the halloysite nanotubes acting as heterogeneous nucleation seeds. RL spectra indicate the presence of $Cr^{3+}$ ions due to impurities from the precursors. However, besides acting as nucleation seeds, the halloysite nanotubes were still able adsorb the $Cr^{3+}$ ions, as evidenced by the decreased RL emission attributed to the $Cr^{3+}$ luminescent centers. Decreased TL intensity upon increasing HNT content together with the RL data suggested that the $Cr^{3+}$ ions have strong participation in the TL emission process as a luminescent center. Surprisingly, the samples with HNT showed up to 6-fold enhanced OSL area intensity and 69-fold OSL initial intensity enhancement, revealing that, by scavenging $Cr^{3+}$, the HNT eliminated a luminescent center that competes with the OSL emission. Therefore, HNTs are promising nanomaterial to enhance the sensitivity of α-$Al_2O_3$ dosimeters with potential application in medical physics, revealing that the presence of HNT in $Al_2O_3$ decreased the density of competing luminescent centers and increased OSL intensity.

## 1. Introduction

The absorbed dose is an essential quantitative parameter used in medical physics, since the interaction of ionizing radiation beams with matter, such as body tissues, causes biological damage in a dose-dependent way (1,2). Therefore, producing more sensitive dosimeters allows for more accurate dosimetric measurements. The optically stimulated luminescence (OSL) and thermoluminescence (TL) consist of dosimetric techniques based on the emission of light by an insulator or semiconductor previously exposed to ionizing radiation and, upon an external stimulus with the light of appropriate wavelength (OSL) or heat (TL) (3–5). Understanding the OSL and TL phenomena is required to analyze the energy levels of the crystal structure of these dosimeters. Initially, when the material is exposed to ionizing radiation, the ionized electron moves to the conduction band and, as a consequence, a hole is created in the valence band. The hole thereby moves through the valence band until it is trapped in a hole trap. The electron in the conduction band can be trapped in electron traps in the materials forbidden band (band gap). External stimulation causes the electron to escape from the trap, which can then recombine with a hole and produce light emission whose intensity is proportional to the absorbed dose.

The OSL measurement can be obtained with the CW-OSL (Continuous Wave) stimulation mode, which consists of continuously illuminating the detector during the readout, resulting in the typical OSL curve characterized by an exponential decay of the light intensity as a function of the stimulation time (3,6,7). The initial OSL intensity is caused by many electron/hole recombinations. It decreases as a function of time because the trapped charge concentration reduces as the electron escapes from the traps (3,8). Both the initial OSL intensity and the area under the OSL curve are proportional to the absorbed dose (3).

Similarly, the radioluminescence (RL) phenomenon could reveal some information about the material's properties, which is a prerequisite for radiation dosimetry applications. The RL emission, as opposed to OSL (9), is not produced by electron trapping, but by the immediate electron-hole recombination that occurs while the sample is irradiated by ionizing radiation beams. Thus, it is perceptible that if electron trapping occurs during the radioluminescence measurement of a sample, the RL intensity will be reduced. On the other hand, an increase in RL intensity indicates that there are more luminescent centers in the sample. Therefore, radioluminescence can identify the luminescent center that exists in a specific sample.

Historically, many studies analyzed the OSL properties in the materials. As an example we can cite the oxides such as beryllium oxide (BeO) (10) and aluminum oxide ($Al_2O_3$) (7,11), doped halides such as KCl: Eu (12), KBr: Eu, NaCl: Cu, and BaFBr: Eu, sulfates as $CaSO_4$: Dy (13), and sulfides as

MgS: Eu (13,14) and CaS:Eu (13). Among these materials, aluminum oxide has the highest dosimetric sensitivity (7). However, the synthesis of aluminum oxide is a major problem, due to the impossibility to control the amount of carbon that dopes these materials during calcination (10,11). Besides, its elevated cost leads to commercial restrictions. Aiming to produce more sensitive and compact aluminum oxide dosimeters, we employed halloysite nanocrystals (HNTs) during the synthesis of aluminum oxide.

Halloysite nanotubes (HNTs) belong to the kaolin group and are compounded by aluminosilicate. These nanoclays have a natural hollow and coiled tubular structure (15–17). Furthermore, the multilayered walls of these nanoclays are organized by a negatively charged outer surface, due to the Si-OH group, and a positively charged inner surface, caused by Al-OH groups (18–20). In virtue of this bivalent spatial separation, halloysite is widely used in a variety of applications, including waste-water treatment (21,22), drug delivery (23–25), and fabrication of nanoeletronics (26,27).

In this work, HNTs were employed to grow the aluminum oxide crystals by heterogeneous nucleation, i.e, the HNTs could act as seeds to promote better $Al_2O_3$ crystals and improve its sensitivity to ionizing radiation. The $Al_2O_3$/HNT composites were characterized by X-ray diffraction and scanning electron microscopy and their dosimetric properties were evaluated by RL, TL and OSL.

## 2. Materials and methods

Aluminum nitrate nonahydrate ($Al(NO_3)_3 \cdot 9H_2O$, 98.00%) from Synth, urea ($NH_2CONH_2$, 99.45%) from NEON, and halloysite nanoclay ($H_4Al_2O_9Si_2 \cdot 2H_2O$) (Mw = 294.19 g/mol) obtained from Sigma-Aldrich.

### 2.1 Combustion

Aluminum oxide was produced by the combustion method. For this synthesis, the aluminum nitrate salt $Al(NO_3)_3$ (10 g), employed as cation precursor, urea (2 g) used as fuel, and the mass percentage of halloysite as seeds (0%, 8.80%, 17.60%, 35.20%, and 52.80%) – respective to the mass aluminum oxide powder produced, were mixed in 10 ml of ultrapure water. The solution was transferred to a cylindrical alumina crucible and introduced into a muffle furnace at 500°C for 5 min. The obtained powder was ground and calcinated at 1200°C for 12 h.

### 2.2 Irradiation and OSL measurements

The samples (20 mg) were irradiated with 10 Gy using an X-ray tube (Magnum, Montex Inc.) operating with 48 kVp and 0.2 mA. The OSL curves were collected upon blue (470 nm), using Hoya U340 filters (transmission between 270 nm and 370 nm), with the continuous reading mode (CW-OSL).

### 2.3 Radioluminescence (RL)

RL measurements were obtained with a 48 kVp X-ray tube (Magnum Moxtek, model TUB00045-1) and 100μA tube current. The RL detection system used an optical fiber with an f/2 fused silica lens to capture the sample's emitted light and a CCD spectrometer. RL spectrum of the samples (40 mg) was recorded with a dose rate of 9 Gy/min.

### 2.4 X-ray diffraction (XRD)

The XRD patterns were recorded using a Bruker D8-Discover, with a copper cathode and germanium monochromator. All the XDR diffractograms were carried out in the $2\theta = 16°–99°$ range and $0.7°$ angular step.

### 2.5 Scanning electronic microscopy (SEM)

The samples' images were obtained with the scanning electronic microscopy (SEM) analysis using a Superscan SS-550 (Shimadzu) at 20 kV accelerating voltage and 2 minutes scan time in photo scan mode. A thin gold coating (≈ 20 Å) was applied to the sample using Coater-Balzers SCD 050.

## 3. Results

### 3.1 X-ray powder diffraction and scanning electron microscopy images

To characterize the crystallographic phase of aluminum oxide samples, XRD patterns of pure calcined halloysite and samples of $Al_2O_3$ produced with different mass percentages of halloysite were acquired. According to Figure 1, the aluminum oxide samples presented the most intense peak at $2\theta$ angles of 25.57°, 35.14°, 37.76°, 43.34°, 52.53°, 57.48°, 66.50°, and 68.18° which corresponded to the (1 1 2), (1 1 4), (2 1 0), (2 1 3), (2 2 4), (2 1 6), (3 1 4) and (3 0 0) planes of the hexagonal structure of α-phase aluminum oxide. The α- $Al_2O_3$ pertains to space group R3c (28), with the hexagonal unit cell, which is composed of aluminum planes interleaved by oxygen planes and both had hexagonal close-packed planes. In addition, the analysis of the XRD showed the following lattice parameters (a = b = 7.76012 Å and c = 12.9935 Å). Furthermore, the calcined halloysite showed high-intensity peaks located at $2\theta$ angles of 16.40° (1 1 0), 19.30° (1 0 1), 35.21° (1 1 2), and 40.81° (1 2 2), which are also observed in the sample of Al$_2$O$_3$/HNT with increased HNT mass percentage. The average particle sizes were calculated using the Debye-Scherrer equation and related to the (1 1 4) aluminum oxide plane. The estimated crystallites sizes were 37.2 ± 3.2 nm, 30.6 ± 2.9 nm, and 29.2 ± 2.9 nm for the samples with 0%, 17.60%, and 52.80% of halloysite, respectively. The decreased crystallite size upon increasing the HNT mass percentage suggests that the halloysite may be acting as a nucleation seed.

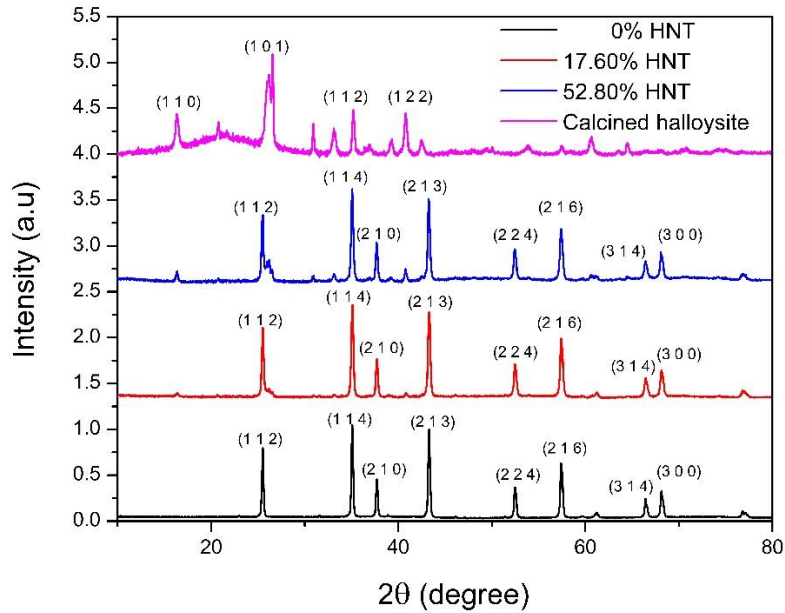

Figure 1: XRD patterns of HNT and samples of aluminum oxide with different mass percentages of HNT, all calcined at 1200 °C during 12h.

SEM images show the surface morphology of the samples at 20K magnification (Figure 2). The calcined halloysite does not exhibit a normally expected tubular structure (24,29). Instead, it appears to have cluster formation that generates structural irregularities. This morphologic difference is probably due to the high calcination temperature, which causes halloysite agglomeration. Additionally, for each aluminum oxide sample, it was noted that the highest the HNT mass percentage, the smoothest the surface. This result is in agreement with the obtained by XRD, both indicating the formation of crystallites with a gradual decrease size upon increasing the HNT content. Therefore, SEM results suggest once again that the halloysite nanoparticles are acting as crystallization seeds, thereby promoting a heterogeneous nucleation process of the aluminum oxide.

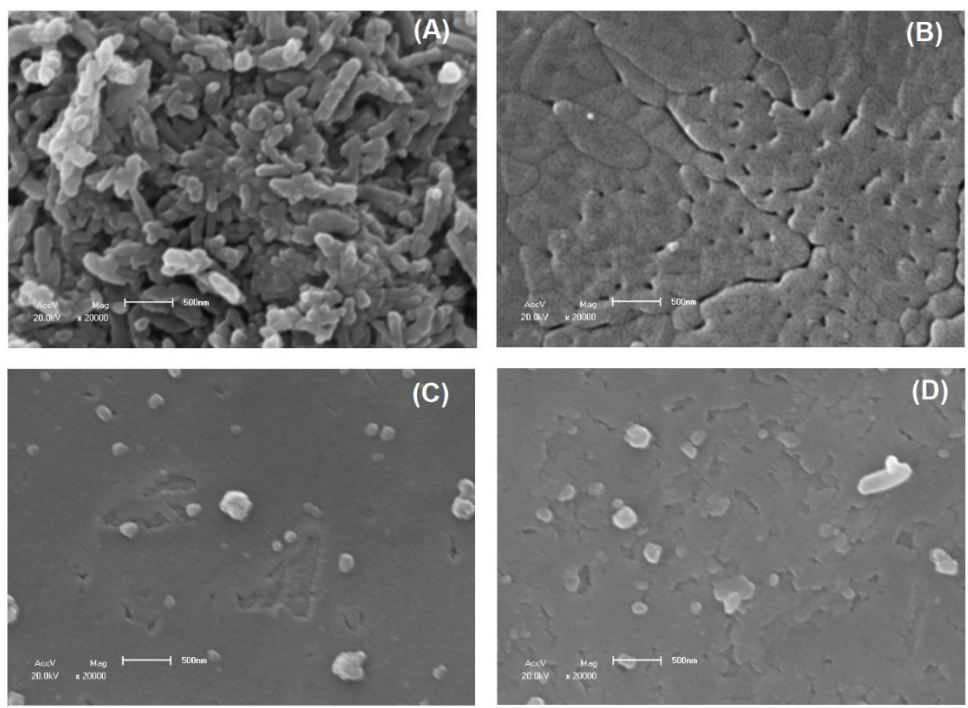

Figure 2: SEM images of (A) calcined halloysite, and (B, C, D) aluminum oxide samples with 0%, 17.60%, and 52.80% of halloysite, respectively.

### 3.2 Radioluminescence (RL)

To analyze the defect of luminescence centers in the samples, Figure 3 shows the average RL spectra and average RL area for varying mass percentages. The maximum RL emission for $Al_2O_3$ dosimeters appears around 420 nm (30–32), but can be shifted when impurities such as transition metals: Fe, Ti, and Cr are added to the dosimeter, increasing the wavelength corresponding to this peak. From Figure 3 (A), it is possible to observe that the RL intensity appears at 694 nm, which is the characteristic emission of $Cr^{3+}$ (33–35) in $Al_2O_3$ matrix. However, during the production of the samples, any $Cr^{3+}$ ion was intentionally added. Thus, a plausible explanation is the small amount of heavy metal ions impurities, about 0.001% present in the aluminum nitrate precursor. Nevertheless, a strong relationship between $Cr^{3+}$ RL emissions in pure alumina crystals has been reported by some authors (36,37). Figure 3 (B) shows a decrease in the RL area as the amount of HNT increases in the samples, suggesting that the halloysite could adsorb the $Cr^{3+}$ ions, removing them from the $Al_2O_3$ matrix. In fact, according to the literature, halloysite nanoclays are extensively used due to their ability to adsorb heavy metal ions (38–40).

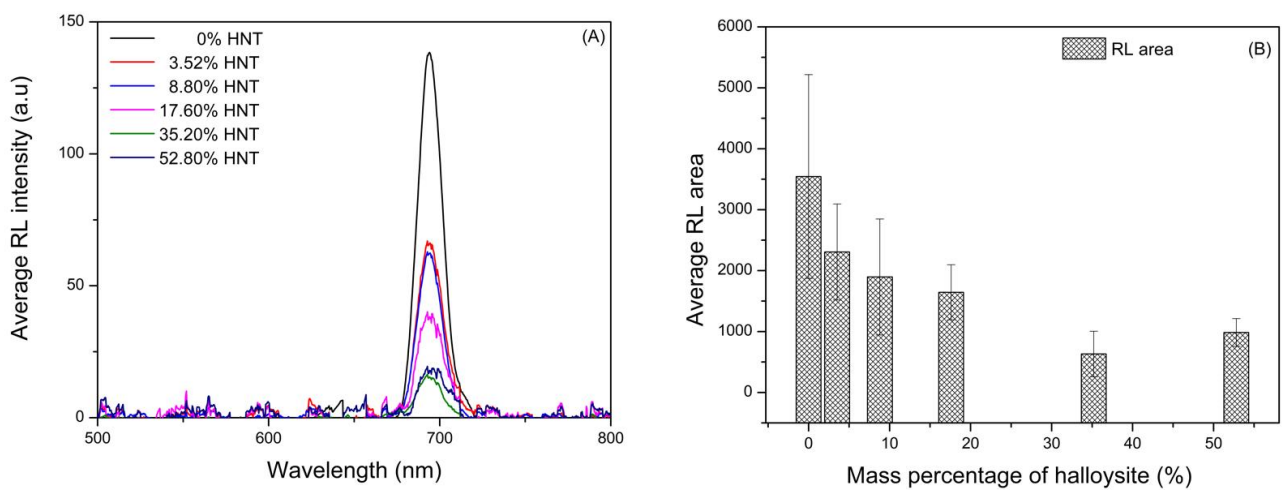

Figure 3: (A) Average RL spectra and (B) Average RL spectra area of $Al_2O_3$ samples with different HNT quantities.

### 3.3 Thermoluminescence (TL)

Thermoluminescence curves from aluminum oxide samples with different HNT quantities were recorded in the 50°C – 400°C temperature range, using a constant heating rate of 5°C/s (Figure 4 A). From Figure 4 (A), the highest TL intensity among all samples was observed around 240°C for the sample with 0% of halloysite (pure $Al_2O_3$). Deconvolution of the pure $Al_2O_3$ glow curve (Figure 4 B) reveals TL peaks around 145°C and 237°C, the last being the main and most intense dosimetric peak. For the pure calcined halloysite, deconvolution of the glow curves (Figure 4 C) reveals TL peaks around 173°C, 270°C, and 298 °C. Finally, Figure 4 D shows TL peaks around 150°C and 209°C for sample with 35.20% HNT. Compared with the pure $Al_2O_3$, the higher temperature peak intensity reduces and shift toward lower temperatures as the mass percentage of halloysite increases. It is worth noting that both pure samples present higher TL intensities compared to the composites, indicating that the presence of HNT in $Al_2O_3$ decreases the density of traps and/or luminescent centers. Decreased number of electronic traps might be related to the smaller crystallite sizes, as revealed by the XRD and SEM results. However, considering the RL results, it is more likely that the $Cr^{3+}$ ions have strong participation in the TL emission process as luminescent centers, once the quenched $Cr^{3+}$ emission upon increasing halloysite mass percentage results in lower RL and TL intensities.

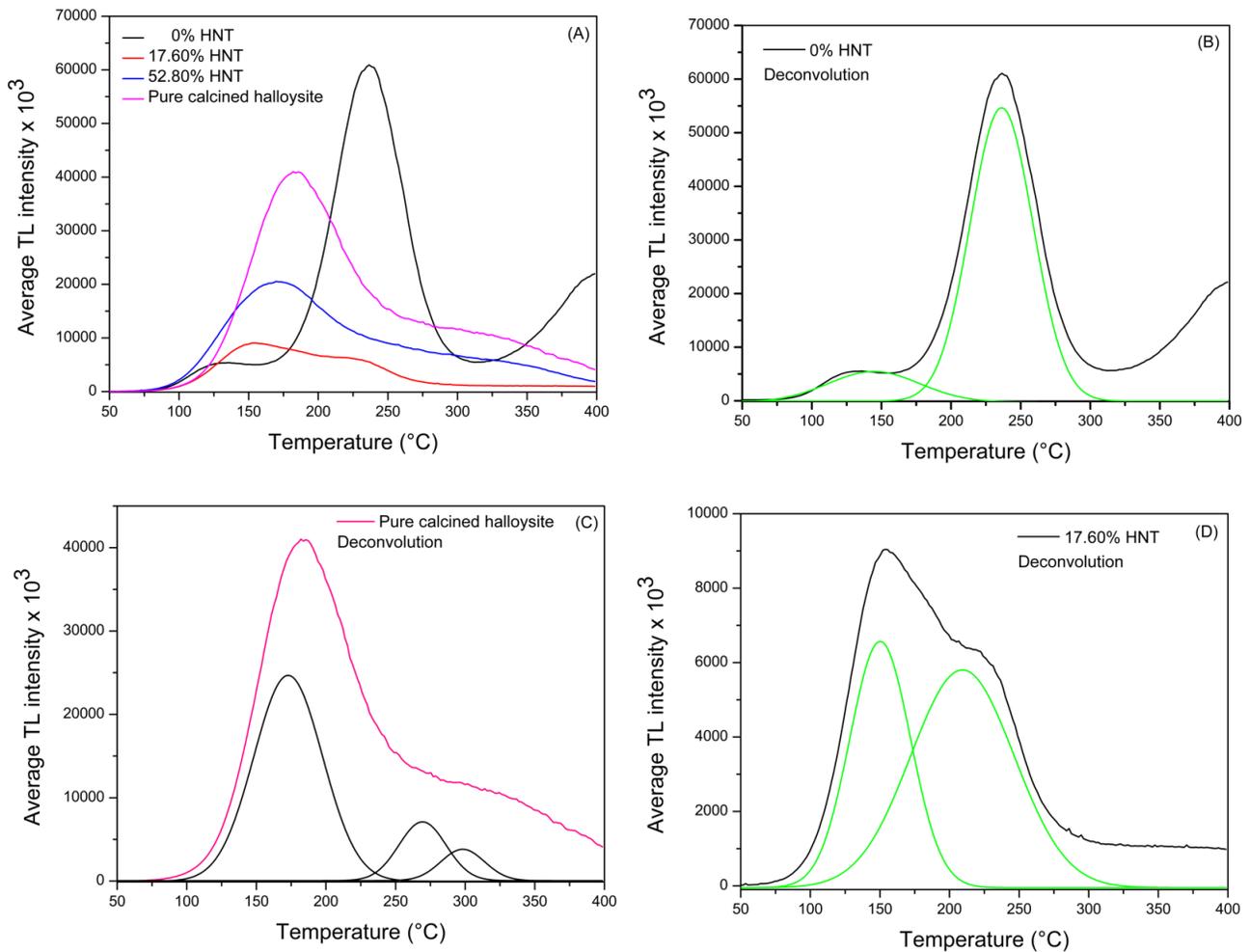

Figure 4: (A) TL glow curves of aluminum oxide samples with varying halloysite mass percentages and irradiated with 100 Gy and (B), (C) and (D) Deconvolution curves for 0% of HNT, pure calcined halloysite samples and 17.60% of HNT, respectively.

### 3.4 Optically stimulated luminescence (OSL)

The OSL curves obtained upon blue stimulation revealed a notorious OSL enhancement for samples containing HNT (Figure 5 (A)). The shape of the OSL curve is the same for most samples -indicating a typical exponential decay due to dosimetric trap concentration (7).

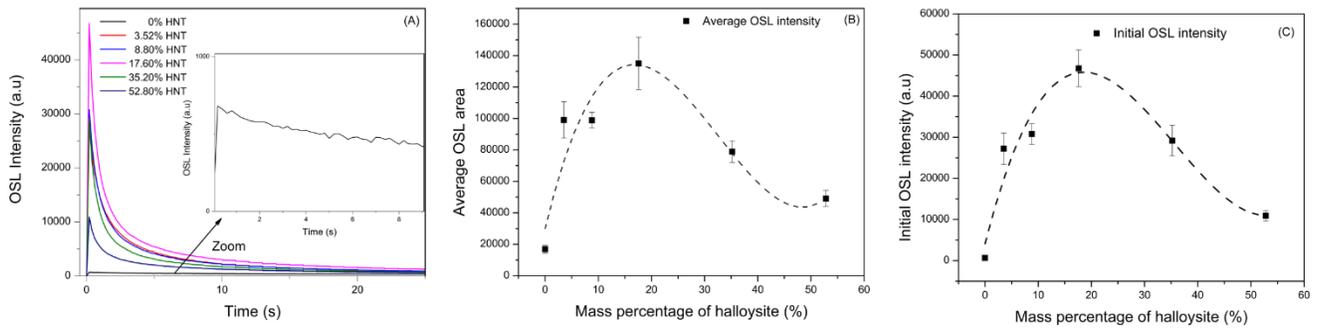

Figure 5: Samples of aluminum oxide with 20 mg and irradiated by 10 Gy and stimulated with blue light. (A) CW-OSL decay curve. The inset corresponds to a zoom in the 0% HNT OSL curve. (B), (C) Average OSL area and initial OSL intensity with different quantities of HNT, respectively.

To investigate the OSL enhancement, Figure 5 (B) shows the OSL intensity, in terms of the integral area, as a function of the HNT mass percentage. It can be noted that an enhanced OSL intensity for samples of $Al_2O_3$ produced in presence of halloysite, with maximum enhancement around 6-fold for samples synthesized with 17.60% HNT mass percentage. This result suggests that either the observed OSL increase could be due to the presence of halloysite nanocrystals itself or HNTs acts as seeds, promoting the growth of higher quality aluminum oxide. Furthermore, Figure 6 (B) shows an enhancement around 69-fold over the ratio of samples with 17.60% and 0% of halloysite.

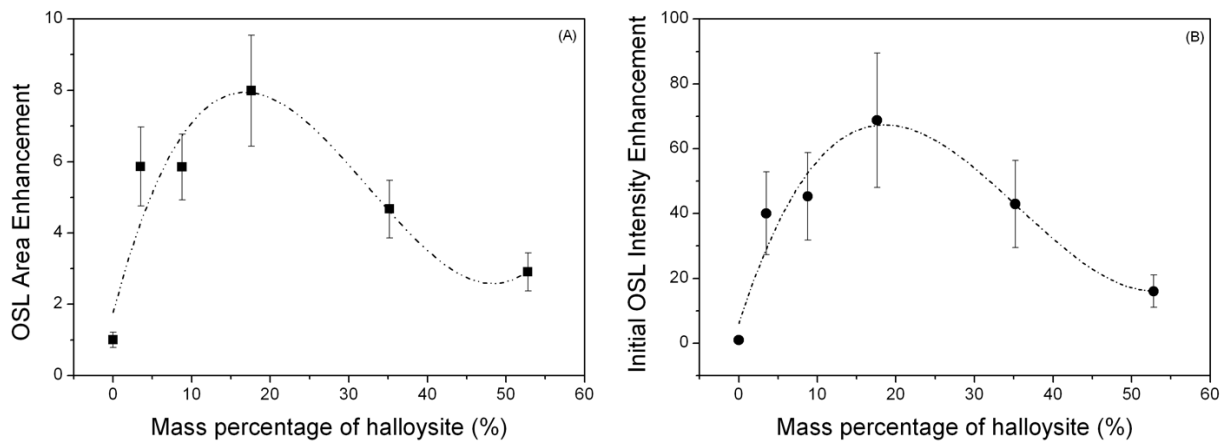

Figure 6: OSL Enhancement (A) OSL Area (B) Initial OSL Intensity.

Figure 7 shows the OSL intensities for the samples with 0% and 17.60% of halloysite, and for the pure HNT in both calcined and non-calcined conditions. The results from Figure 6 revealed that the OSL intensity from pure halloysite, either calcined or non-calcined, is much lower than the OSL intensity of the composites, indicating that the enhanced OSL is not caused by the OSL emitted by the halloysite nanocrystals.

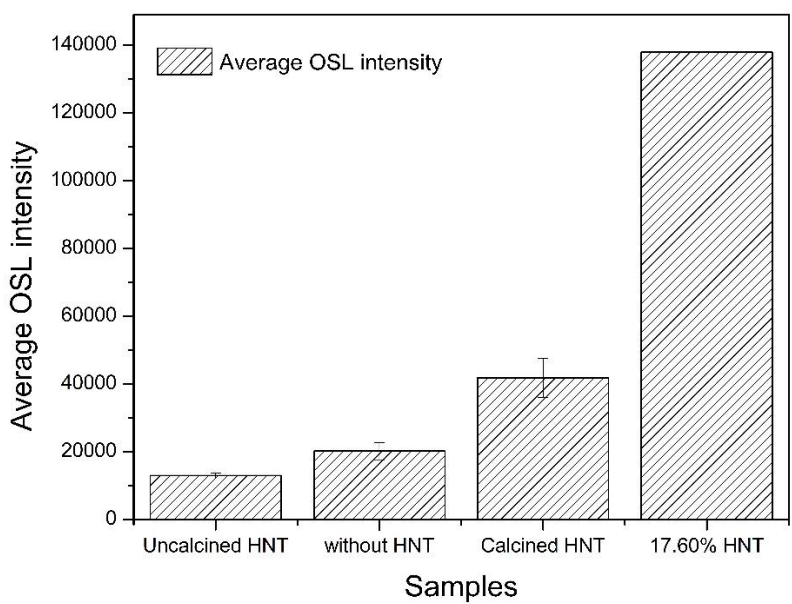

Figure 7: Average OSL intensities for 20 mg samples irradiated with 10 Gy.

Here, it should be noted that OSL emission is collected only in the 270-390 nm wavelength, whereas TL glow curves are collected in the whole visible range (~300-700nm) wavelengths. Therefore, considering that the TL glow curves suggested a decreased density of electronic traps and/or luminescence centers after the addition of halloysite during the synthesis of aluminum oxide, the enhanced OSL intensity may be a consequence of a decreased $Cr^{3+}$ RL emission at 694 nm upon increasing the halloysite mass percentage, as suggested by RL spectra. In other words, the OSL readout system is not able to detect $Cr^{3+}$ luminescence, which will act as a competing luminescence center. Furthermore, it is noted that there is an optimal amount of halloysite to increase the OSL intensity. This could be related to the ability of the HNT to scavenge heavy metals. Upon increasing the HNT concentration, the HNT nanoparticles can agglomerate (41,42), decreasing the effective surface area, thereby reducing the ability to adsorb and remove $Cr^{3+}$ ions from the $Al_2O_3$ matrix. This hypothesis is reinforced by the fact that RL intensity also reached a plateau for HNT mass percentage above 35.20%. In this sense, the decreased density of trap states, as suggested by the TL glow curves, may be related to the decreased OSL for HNT mass percentage $\geq$ 35.20%.

**Conclusion**

In this work, the X-ray diffraction results showed that both samples of aluminum oxide with halloysite and without halloysite presented the α crystallographic phase, which is the most stable phase of aluminum oxide. Furthermore, it was noted that as the amount of halloysite increased in the samples, the peaks corresponding to angles 16.40°, 19.30°, 52.81°, and 40.81° also increased. Besides, the SEM revealed that the HNTs area acting as a nucleation seeds. Moreover, the RL investigations showed the presence of $Cr^{3+}$ ions, which reduces as the mass percentage of halloysite increases in the samples, indicating the heavy metal adsorption ability of this nanotube. Subsequently, the TL curve exhibited a decrease upon increasing quantity of halloysite in the samples further suggesting that the halloysite was adsorbing the $Cr^{3+}$ ions. In addition, the OSL area and initial OSL intensity revealed an enhancement (about 6-fold and 69-fold), respectively for aluminum oxide samples with 17.60%. Therefore, this work shed new light on the production of OSL dosimetric materials synthesized with aluminum oxide and halloysite, due to the finding of halloysite can act as a seed providing a surface for this crystal to grow.

**Acknowledgments**

The authors gratefully acknowledge the financial support of the Brazilian funding agencies CNPQ, CAPES, and FAPESP (21/00786-0); Eldereis de Paula and Cassiana Viccari Sacilotto for technical assistance, Rodrigo Ferreira Silva for the SEM images, and Centro de Instrumentação, Dosimetria e Radioproteção (CIDRA) for the TL measurements.

**Funding**

This work was funded by the Brazilian funding agencies CNPQ, CAPES, and FAPESP (21/00786-0).

**Conflict of Interest**

The authors declare that they have no known competing financial interests or personal relationships that could have appeared to influence the work reported in this paper.

**References**

1. Clements BW, Casani JAP. Nuclear and Radiological Disasters. Disasters Public Heal. 2016;357–83.

2. Johns HE, Cunningham JR. The Physics of radiology, Edition 4. Charles River. 1983.

3. Yukihara EG, McKeever SWS. Optically Stimulated Luminescence. Optically Stimulated Luminescence. 2011.

4. Bhatt BC. Thermoluminescence, optically stimulated luminescence and radiophotoluminescence dosimetry: an overall. Radiat Prot Env. 2011;34:6–16.

5. Mckeever SWS. Thermoluminescence of Solids. University Press; 1985.

6. Sahil, Kumar R, Yadav MK, Kumar P. OSL and TA-OSL properties of Li2B4O7:Al for radiation dosimetry. J Alloys Compd [Internet]. 2022;908:164628. Available from: https://doi.org/10.1016/j.jallcom.2022.164628

7. Yukihara EG, McKeever SW. Optically stimulated luminescence (OSL) dosimetry in medicine. Phys Med Biol. 2008;53(20).

8. Jain A, Seth P, Tripathi A, Kumar P, Aggarwal S. TL/OSL response of carbon ion beam irradiated NaMgF3:Tb. J Lumin. 2020;222(December 2019).

9. Polf JC, Yukihara EG, Akselrod MS, McKeever SWS. Real-time luminescence from Al2O3 fiber dosimeters. Radiat Meas. 2004;38(2):227–40.

10. Watanabe S, Gundu Rao TK, Page PS, Bhatt BC. TL, OSL and ESR studies on beryllium oxide.

J Lumin [Internet]. 2010;130(11):2146–52. Available from: http://dx.doi.org/10.1016/j.jlumin.2010.06.009

11. Akselrod MS, Bøtter-Jensen L, McKeever SWS. Optically stimulated luminescence and its use in medical dosimetry. Radiat Meas. 2006;41(SUPPL. 1):78–99.

12. Nanto, H.; Murayama, K.; Usuda, T.; Taniguchi, S.; Takeuchi N. Optically stimulated luminescence in KCl:Eu single crystals. 1993;47 (1-4),:281–4.

13. Bøtter-Jensen L, McKeever SWS, Wintle AG. Optically Stimulated Luminescence Dosimetry. Optically Stimulated Luminescence Dosimetry. 2003.

14. Loncke F, Vrielinck H, Matthys P, Callens F, Tahon JP, Leblans P. Paramagnetic Eu2+ center as a probe for the sensitivity of CsBr x-ray needle image plates. Appl Phys Lett. 2008;92(20).

15. Satish S, Tharmavaram M, Rawtani D. Halloysite nanotubes as a nature's boon for biomedical applications. BJGP Open. 2019;6:1–16.

16. Yurdacan HM, Murat Sari M. Functional green-based nanomaterials towards sustainable carbon capture and sequestration [Internet]. Sustainable Materials for Transitional and Alternative Energy. INC; 2021. 125–177 p. Available from: http://dx.doi.org/10.1016/B978-0-12-824379-4.00004-5

17. Yuan P, Tan D, Annabi-Bergaya F. Properties and applications of halloysite nanotubes: Recent research advances and future prospects. Appl Clay Sci [Internet]. 2015;112–113:75–93. Available from: http://dx.doi.org/10.1016/j.clay.2015.05.001

18. Nuruzzaman M, Liu Y, Rahman MM, Dharmarajan R, Duan L, Uddin AFMJ, et al. Nanobiopesticides: Composition and preparation methods [Internet]. Nano-Biopesticides Today and Future Perspectives. Elsevier Inc.; 2019. 69–131 p. Available from: https://doi.org/10.1016/B978-0-12-815829-6.00004-8

19. Stein R. Clay minerals. Encycl Earth Sci Ser. 2016;Part 2(Figure 1):87–93.

20. Abdullayev E, Lvov Y. Halloysite for Controllable Loading and Release [Internet]. 1st ed. Vol. 7, Developments in Clay Science. Elsevier Ltd.; 2016. 554–605 p. Available from: http://dx.doi.org/10.1016/B978-0-08-100293-3.00022-4

21. Grylewicz A, Mozia S. Polymeric mixed-matrix membranes modified with halloysite nanotubes for water and wastewater treatment: A review. Sep Purif Technol [Internet]. 2021;256(July 2020):117827. Available from: https://doi.org/10.1016/j.seppur.2020.117827

22. Yu L, Wang H, Zhang Y, Zhang B, Liu J. Recent advances in halloysite nanotube derived composites for water treatment. Environ Sci Nano [Internet]. 2016;3(1):28–44. Available from: http://dx.doi.org/10.1039/C5EN00149H

23. Liu H, Wang ZG, Liu SL, Yao X, Chen Y, Shen S, et al. Intracellular pathway of halloysite nanotubes: potential application for antitumor drug delivery. J Mater Sci. 2019;54(1):693–704.

24. Fizir M, Dramou P, Dahiru NS, Ruya W, Huang T, He H. Halloysite nanotubes in analytical sciences and in drug delivery: A review. Microchim Acta. 2018;185(8).

25. Guo M, Wang A, Muhammad F, Qi W, Ren H, Guo Y, et al. Halloysite nanotubes, a multifunctional nanovehicle for anticancer drug delivery. Chinese J Chem. 2012;30(9):2115–20.

26. Zhang Y, Tang A, Yang H, Ouyang J. Applications and interfaces of halloysite nanocomposites. Appl Clay Sci [Internet]. 2016;119:8–17. Available from: http://dx.doi.org/10.1016/j.clay.2015.06.034

27. Lvov Y, Wang W, Zhang L, Fakhrullin R. Halloysite Clay Nanotubes for Loading and Sustained Release of Functional Compounds. Adv Mater. 2016;28(6):1227–50.

28. Ramogayana B, Santos-Carballal D, Maenetja KP, De Leeuw NH, Ngoepe PE. Density Functional Theory Study of Ethylene Carbonate Adsorption on the (0001) Surface of Aluminum Oxide α-$Al_2O_3$. ACS Omega. 2021;6(44):29577–87.

29. Yu D, Wang J, Hu W, Guo R. Preparation and controlled release behavior of halloysite/2-mercaptobenzothiazole nanocomposite with calcined halloysite as nanocontainer. Mater Des. 2017;129(March):103–10.

30. Nyirenda AN, Chithambo ML. Spectral study of radioluminescence in carbon-doped aluminium oxide. Radiat Meas [Internet]. 2018;120(November 2017):89–95. Available from: https://doi.org/10.1016/j.radmeas.2018.06.026

31. Rodriguez MG, Denis G, Akselrod MS, Underwood TH, Yukihara EG. Thermoluminescence, optically stimulated luminescence and radioluminescence properties of $Al_2O_3$:C,Mg. Radiat Meas [Internet]. 2011;46(12):1469–73. Available from: http://dx.doi.org/10.1016/j.radmeas.2011.04.026

32. Jursinic PA. Changes in optically stimulated luminescent dosimeter (OSLD) dosimetric characteristics with accumulated dose. Med Phys. 2010;37(1):132–40.

33. Lewis PM, N K, Hebbar N D, Choudhari KS, Kulkarni SD. $Cr^{3+}$ doped $Al_2O_3$ nanoparticles:

Effect of Cr3+ content in intensifying red emission. Curr Appl Phys [Internet]. 2021;32(October 2021):71–7. Available from: https://doi.org/10.1016/j.cap.2021.10.003

34. Du Y, Cai WL, Mo CM, Chen J, Zhang LD, Zhu XG. Preparation and photoluminescence of alumina membranes with ordered pore arrays. Appl Phys Lett. 1999;74(20):2951–3.

35. da Cunha GC, Abreu CM, Peixoto JA, Romão LPC, Macedo ZS. A Novel Method For Fabricating Cr-Doped Alpha-Al2O3 Nanoparticles: Green Approach To Nanotechnology. J Inorg Organomet Polym Mater. 2017;27(3):674–84.

36. Zuo C, Jagodzinski PW. R-line luminescence from trace amounts of Cr3+ in aluminum oxide and its dependence on sample hydration. Appl Spectrosc. 2002;56(8):1055–8.

37. Kristianpoller N, Rehavi A, Shmilevich A, Weiss D, Chen R. Radiation effects in pure and doped Al2O3 crystals. Nucl Instruments Methods Phys Res Sect B Beam Interact with Mater Atoms. 1998;141(1–4):343–6.

38. Liu X., Tournassat C, Grangeon S, Kalinichev AG, Takahashi Y, Fernandes MM. Molecular-level understanding of metal ion retention in clay-rich materials. Nat Rev Earth Environ. 2022;3(7):461–76.

39. Hermawan AA, Chang JW, Pasbakhsh P, Hart F, Talei A. Halloysite nanotubes as a fine grained material for heavy metal ions removal in tropical biofiltration systems. Appl Clay Sci [Internet]. 2018;160(September 2017):106–15. Available from: https://doi.org/10.1016/j.clay.2017.12.051

40. Kiani G. High removal capacity of silver ions from aqueous solution onto Halloysite nanotubes. Appl Clay Sci [Internet]. 2014;90:159–64. Available from: http://dx.doi.org/10.1016/j.clay.2014.01.010

41. Tas CE, Ozbulut EBS, Ceven OF, Tas BA, Unal S, Unal H. Purification and Sorting of Halloysite Nanotubes into Homogeneous, Agglomeration-Free Fractions by Polydopamine Functionalization. ACS Omega. 2020;5(29):17962–72.

42. Pasbakhsh P, De Silva R, Vahedi V, Jock Churchman G. Halloysite nanotubes: prospects and challenges of their use as additives and carriers – A focused review. Clay Miner. 2016;51(3):479–87.